\documentclass[letterpaper,twocolumn,10pt]{article}
\usepackage{usenix}
\usepackage{filecontents}
\usepackage[font=small]{caption}
\usepackage[T1]{fontenc}
\usepackage[available]{usenixbadges}
\usepackage{tikz}
\usepackage{amsmath}
\usepackage[table,xcdraw]{xcolor}
\usepackage[dvipsnames]{xcolor}
\definecolor{myyellow}{HTML}{E1BE6A}
\definecolor{mygreen}{HTML}{40B0A6}
\usepackage[normalem]{ulem}
\useunder{\uline}{\ul}{}
\usepackage{longtable}
\usepackage{tabularx}
\usepackage{xurl}

\usepackage{filecontents}
\usepackage{amsmath,amssymb,amsfonts}
\usepackage{algorithmic}
\usepackage{graphicx}
\usepackage{textcomp}
\usepackage{booktabs}
\usepackage{dcolumn}
\usepackage[flushleft]{threeparttable}
\usepackage{comment}
\newcolumntype{?}{!{\vrule width 1pt}}

\usepackage{hyperref}

\definecolor{amethyst}{rgb}{0.6, 0.4, 0.8}

\newcommand{\totalStandards}{12,201}
\newcommand{\securityStandards}{3,920}
\newcommand{\percentSecurity}{32.1}
\newcommand{\hasExamplesSecurity}{2,841}
\newcommand{\hasExamplesStates}{31}
\newcommand{\securityCategoryStandards}{1,669}
\newcommand{\securityCategoryStandardsPercent}{13.7} 
 
\newcommand{\elementarySecurityStandards}{1,174}
\newcommand{\middleSecurityStandards}{859}
\newcommand{\highSecurityStandards}{1,670}
\newcommand{\numCodes}{109}
\newcommand{\numSecStates}{29}
\newcommand{\percentOfSecurityCalledSecurity}{42.6} 
\newcommand{\spstandards}{S\&P standards}

\newcommand{\sap}{security and privacy}
\newcommand{\snp}{S\&P}

\newcommand{\codes}{topics}
\newcommand{\teachingstandards}{K-12 teaching standards}
\newcommand{\tech}{future technology professionals}
\newcommand{\nontech}{general internet/tech users}
\newcommand{\standard}[2]{\textit{#1} (#2)}
\newcommand{\standardP}[2]{\textit{#1}, #2}
\renewcommand{\quote}[1]{``\textit{#1}''}
\newcommand{\quotepar}[2]{``\textit{#1}'' (#2)}
\newcommand{\summary}[1]{\vspace{.1cm} \noindent \textbf{#1}}
\begin{document}

\title{\Large \bf Characterizing Security and Privacy Teaching Standards for Schools in the United States}

\def\emptyauthor{Anonymized for Submission}

\author{
{\rm Katherine Limes}\\
Colorado School of Mines
\and
{\rm Nathan Malkin}\\
New Jersey Institute of Technology
\and
{\rm Kelsey R.\ Fulton}\\
Colorado School of Mines
}





\maketitle
\pagestyle{empty}

\begin{abstract}
Increasingly, students begin learning aspects of security and privacy during their primary and secondary education (grades K-12 in the United States). Individual U.S.\ states and some national organizations publish teaching standards---guidance that outlines expectations for what students should learn. However, research has not yet examined what is covered by these standards and whether the topics align with what the broader security and privacy community thinks students should know. To shed light on these questions, we manually examined 12,201 standards from all U.S. states and eight national organizations. We labeled 3,920 of them as being related to security and privacy, further classifying these into 109 topics ranging from encryption to appropriate online behavior. 
\end{abstract}

\section{Introduction}
\label{sec:intro}

Education about security and privacy (\snp{}) is critical, whether the goal is to create future security professionals, train developers to prioritize security and privacy, or simply equip users to protect themselves online.
Increasingly, such education starts before university:
many primary and secondary schools have added classes on computer science and technology,
which often cover \sap{} topics.
But what exactly are they teaching?
What skills and topics are deemed most important, and which are emphasized less?
Is content tailored to future tech professionals or informed end users?

These questions are difficult to answer.
The United States, for example, has over 120,000 schools~\cite{NationalCenterForEducationStatistics22},
and the majority of school districts offer at least some computer science education~\cite{Code.org22}.
It is unclear to what extent any schools' curricula can be extrapolated to others, as
teachers often have substantial discretion in what they teach.
Yet, there are some common denominators.
In classes meant to prepare students for standardized tests (like Advanced Placement and International Baccalaureate exams), teachers must teach the topics that will be tested.
Besides this, curriculum development is often guided by state teaching standards.

\begin{figure*}
    \centering
    \includegraphics[width=1\linewidth]{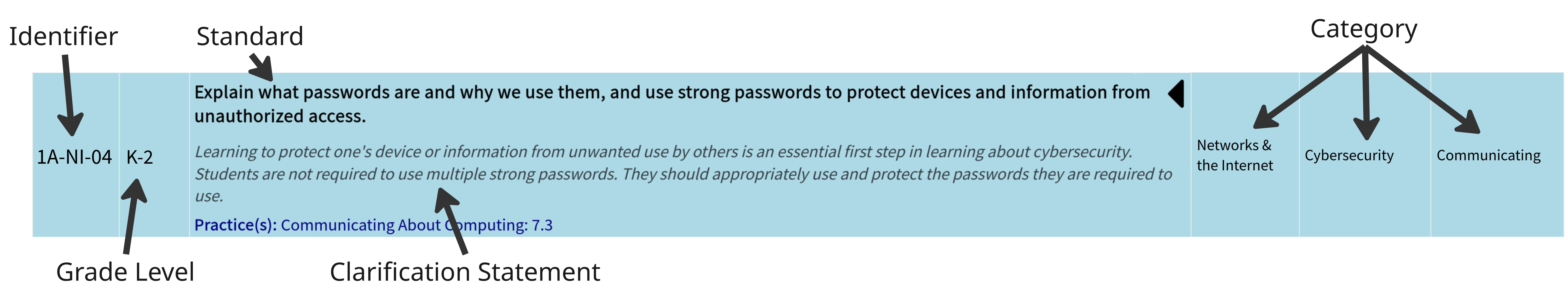}
    \caption{Example CS teaching standard}
    \label{fig:standard}
\end{figure*}

\emph{Teaching standards}
are documents that specify knowledge and performance expectations for students studying a particular subject
at a given grade level.
(See example standard in Figure~\ref{fig:standard}.)
While they define topics that should be taught,
they do not provide detailed guidance about \emph{how} subjects should be covered.
Nonetheless, because most teachers are generally expected to comply with standards adopted by their state,
teaching standards provide a point of consistency between schools and curricula.
They therefore offer an interesting vantage point for understanding what states want their students to learn.
They are also a potentially powerful mechanism of effecting change:
rather than targeting individual curricula at thousands of schools, standards can be improved, and the changes will propagate to individual classrooms.

Our study aims to systematically examine teaching standards for their security and privacy content and begin the process of evaluating their utility and usability, addressing a gap in the literature regarding this topic.
Our vision is that, by characterizing the content and highlighting potential gaps, we can identify a path toward more effective security and privacy education, with greater empirical grounding and larger involvement from the academic and industry communities.
To do this, we formulated the following research questions:

\smallskip
\noindent
\textbf{RQ1.} What security and privacy topics are covered by K-12 computer science teaching standards in the U.S.?\\
\textbf{RQ2.} How are U.S. K-12 security and privacy teaching standards presented (e.g., organization, presence of examples)? \\
\textbf{RQ3.} How do standards differ between states/grade levels?\\

\smallskip
To answer our research questions,
we analyzed computer science teaching standards
from all U.S. states and eight national organizations, yielding \totalStandards. Through manual analysis, we labeled \securityStandards{}~of them as being related to security and privacy. These standards spanned \numCodes{} security and privacy topics ranging from technical subjects, like encryption and network security, to social subjects such as laws, ethics, and appropriate online behavior.
Standards often lacked examples of what to teach. When examples were available, they ranged from detailed clarification statements to short lists of key terms.
Topic coverage varied heavily by state and grade level. Even the most common topics were not prescribed in every state.
Standards tended to focus on fundamental security and privacy topics at the elementary level, and technical topics at the high school level. 

Based on our results,
we recommend clearer separation of standards between \sap{} fundamentals (important for everyone) and more specialized content for future computing professionals, since we found both in the standards.
We also recommend greater investment in teachers and resources for them.


\section{Background and Related Work}
\label{sec:background}

This section provides background on U.S. K-12 education and teaching standards and discusses related work on teaching computer science (CS) and \sap{} (\snp{}) in K-12 schools.

\subsection{Teaching standards}
The U.S. education system is divided into different grade levels, from kindergarten (ages 5--6) to 12 (ages 17--18). These grade levels are often banded into three different groups: elementary (K--5), middle (6--8), and high school (9--12). U.S. states may provide teachers with standards, which are learning objectives for courses at different grade levels. Standards do not explicitly prescribe a curriculum; rather, they indicate what knowledge each student should have at the end of the course.
Each state creates its own standards at the direction of its legislature.
The state puts together a committee of educators and domain experts to draft the standards, releases the draft for public comment, and revises as appropriate. While a state may provide hundreds or thousands of standards, an individual teacher will use only the subset of those standards relevant to their course.
For core subjects (e.g., social studies, math),
teachers are obligated to show that their curriculum meets the learning objectives set by the standards.
For electives, as CS often is,%
\footnote{
    As of 2025, only 12 states have created some form of computer science graduation requirement, only three of those have implemented that requirement, and only one state (North Dakota) required students to learn \sap{} before graduation~\cite{StateComputerScienceEducation2024}. 
}
there may be more flexibility,
and teachers
may choose to add topics beyond the standards or de-emphasize standards that are outdated or not perceived to be useful. If standards are not required, teachers may still use them to decide what their course should cover. They may draw from more than one set of standards for this purpose, or try to align their course with non-CS learning objectives.

\summary{Anatomy of a standard.}
%
We provide an example standard in Figure~\ref{fig:standard}.
Every standard has text, which describes the topic(s) to be covered. 
Additionally, states often provide an identifier, a category and sometimes a subcategory, and a grade level to which the standard applies. Some standards will include examples: additional detail about the intent of the standard or how it should be taught. 

\subsection{Teaching CS in K-12 schools}

Computer science education has been the focus of research for over a decade.
The Institute for Advancing Computing Education has
identified 1,260 papers covering this topic~\cite{McGill25}.
Research in this space focuses on evaluating activities for students, building and evaluating curricula, building and evaluating resources (e.g., tools and evaluation instruments), developing training for teachers, and understanding the current CS education research space~\cite{McGill25,Gardner22,Upadhyaya20,Garneli15}. Some work in this space has specifically focused on K-12 CS teaching standards by evaluating teacher growth using CS teaching standards~\cite{McGill23} and exploring the challenges of implementing standards in New Zealand~\cite{Samarasekara22}. Most closely related to our work is an evaluation of the topics covered by K-12 CS teaching standards, identifying three major concepts: computer literacy, digital citizenship, and computational thinking~\cite{Guo20}. We expand on this by focusing specifically on security and privacy.

\subsection{Teaching \snp{} in K-12 schools}
\label{ss:snp_k12}
Prior work has also focused on teaching \sap{} in schools more broadly, with little to no focus on standards. There has been substantial prior work exploring how to build a number of activities, tools, and interventions to best facilitate security and privacy education~\cite{McGill22, Khan22,Percival22,Luh22,DeBello22,Toledo22,McGill21,Yett20,Zinkus19,Ledeczi19,Amo19,Dunn18,Jin18,Chattopadhyay18,Svabensky18,Mishra17,Jethwani17,Nix14,Rursch13}. While prior work shares our goal of improving \snp{} education, we focus on what students are learning about \sap{} through analysis of \teachingstandards{} rather than how to most effectively teach \sap{}.


Other work has focused on teachers. Exploring gaps in teachers' \sap{} backgrounds and coverage of topics, Hipsky et al.\ identified gaps in teacher \snp{} education, regulatory requirements, and the \sap{} topics covered in schools. They identified a focus on internet and cyber safety in the topics covered~\cite{Hipsky15}. 
Exploring how \sap{} is taught, Kilhoffer et al.\ interviewed high school students and teachers, finding that cybersecurity is often taught in non-technical classes, and teachers employ a number of  strategies to teach these subjects, including holding discussions, technical activities, games, and content creation. This research identified a number of challenges, such as a lack of qualified personnel and personal knowledge barriers~\cite{Kilhoffer23}.

Most relevant to our research is an extended abstract by Kumar et al., describing preliminary work on privacy literacy topics covered in K-12 teaching standards~\cite{Kumar23}.
The researchers report identifying 44 states with privacy standards, generally covering topics related to managing passwords and being careful about posting things online. Our research expands on this work by focusing on security \emph{and} privacy and providing an in-depth comparative analysis of the topics.

\section{Method}
\label{sec:method}
\label{sec:methods}
The goal of our research is to understand what topics current K-12 \sap{} teaching standards cover, how those topics are presented in the standards, and how topic coverage varies across states and grades.
In this section, we describe our collection and analysis of \spstandards{}.

\subsection{Standards analysis}
\label{sec:methods-analysis}
Only Georgia explicitly outlines a security and privacy course in its CS standards.
Students in other states also learn \snp{} topics, but the courses are not dedicated to \snp{}.
We reasoned that computer science and technology classes were most likely to contain this content and therefore focused our analysis on them.
While other classes (e.g., social studies) may cover relevant subjects, our manual analysis methodology precluded us from examining all school subjects in this initial exploratory study.
Future work can build on our findings and perhaps use more automated approaches to expand the coverage.

\summary{Collecting CS standards.}
We included all standards that were available as of June 2024 in our analysis.%
\footnote{As of February 2026, Colorado, Florida, Indiana, Louisiana, North Dakota, South Dakota, and Wyoming have updated or amended their standards since the time of analysis. We did not reanalyze them, but an informal review suggests that only Colorado and Florida made significant changes, and that those changes would not introduce new topics or significantly alter the prevalence of topics within our dataset.}
We searched online for computer science standards from each state, yielding standards for 47 states. All states post their standards on a state-owned website, so finding one grade band easily led to others. For completeness, we cross-referenced our findings with Code.org’s State of CS report~\cite{StateComputerScienceEducation2024}, which independently tracks which states have CS standards. Our search identified three more states with standards than the report indicated.\footnote{Nebraska and South Dakota released standards after the report was published. Minnesota's standards are not intended for a standalone CS class; we included them because they provided examples of how CS content might be covered outside of CS courses.}

In addition to states, we also searched for standards developed by well-known CS teaching organizations. We identified CS standards from the Computer Science Teachers Association (CSTA)~\cite{CSTA17} and the International Society for Technology in Education (ISTE)~\cite{ISTE24}, which had been adopted in ten states and one state, respectively. We also identified \sap{} standards developed by the Cybersecurity and Infrastructure Security Agency (CISA), which had not been adopted directly, but were a set of \snp{}-specific standards~\cite{Cyber.org21}. 
In addition to organizations, we also considered exams and certifications that high school students may complete. 
To this end, we analyzed the topic lists from both CS Advanced Placement exams (APCSA~\cite{APCSA} and APCSP~\cite{APCSP}), both CS International Baccalaureate exams (IBCS and IBCSD~\cite{IB}), and the IC3 GS6 Digital Literacy certification~\cite{IC3}. 

\summary{Identifying \sap{} standards.}
While we had collected all computer science standards, our goal was to focus our analysis only on those related to security and privacy.
To achieve this narrower scope, we followed a two-step process.
First, we used a subset of the standards to create a list of topics that we deemed relevant to \snp{}.
Then, when we coded the remaining standards, any other topics were labeled ``not \snp{}'' (rather than something more specific) to save effort.

We started by using open coding~\cite{Braun12} to label the topics covered in \emph{all} CS standards from four states.
Each team member read through the resulting topics and independently judged whether they may be relevant to security or privacy.
We then met as a group to discuss and reach a consensus.
We observed several distinct categories of topics:
\begin{enumerate}
    \setlength{\itemsep}{0pt}\setlength{\parskip}{0pt}
    
    \item Technical topics involving networks or security.
    \item Topics about impacts of computing on individuals and society.
    \item Computing fundamentals.
\end{enumerate}

Our general philosophy was to include topics that would be broadly relevant to \snp{},
such as those that might be present in a course about security.
To that end, we included societal topics if they had plausible ramifications for security,
such as any ethical and legal considerations;
in contrast, we excluded standards that focused solely on economic impacts of computing.
We excluded most computing fundamentals
(e.g., \standardP{Evaluate efficiency, correctness, and clarity of algorithms}{APCS}), 
because, while security does build on them,
we found their immediate relevance low.
However, we included topics relevant to debugging/troubleshooting, considering them as first steps towards investigating security edge cases.
We emphasize that we do not consider these to be final judgments of \snp{} relevance but a practical heuristic for creating an inclusive sample of standards that may be interesting for security researchers to analyze.


%

With the final list of \snp{} topics as the codebook,
two researchers read and labeled \emph{all} CS standards from the remaining states and organizations.
(If a standard was not security- or privacy-related, it was labeled as such.)
They met after coding all standards from a state to resolve disagreements and update the codebook as needed.
IRR was calculated for all states using Kupper-Hafner agreement, which is used in situations where more than one code can be assigned to an item~\cite{Kupper89}, as was the case for standards, which may cover multiple topics.
The average initial IRR was $.62$
but full agreement was reached on all labels after discussion.
We found that most disagreements were due to inherent vagueness of standards%
\footnote{
e.g., \standard{Research and describe the risks and risk mitigation strategies associated with the utilization and implementation of social media and other digital technology implications.}{Arkansas} 
}
and cases where standards listed many examples, making it easy to miss one or two.

\summary{Labeling \sap{} standards.}
As described above, we labeled each standard with the topic it covered.
%
%
We also collected several other pieces of information.
Since states and organizations often categorized their standards, we identified whether the categories explicitly referenced security and/or privacy
(e.g., ``Safety, Privacy, and Security''), as this illuminates what a teacher might find if they searched for these key terms.

Standards are often very general;
e.g.,
\standard{the student will evaluate current and emerging programming security practices}{Virginia}. 
Consequently, teachers may lack sufficient \sap{} background to effectively implement them~\cite{Kilhoffer23}.
We therefore identified standards that contained
any type of additional detail outside of the standard text.

While some \snp{} subjects are immediately applicable by all students
(e.g., how to use and select passwords)
others are likely to only be used in practice by future technology professionals
(e.g., modes of encryption),
even if the process of learning them could be abstractly beneficial to any student.
To better understand the distribution of topics among these two broad categories,
the team classified
every label in the codebook as being beneficial to \nontech{}, or primarily to future technology professionals, discussing each code until a consensus was reached.
This distinction was used purely as a lens to understand the target audience
(we report our findings in \S\ref{sec:results-audience});
all topics were included equally in our analysis.

\subsection{Limitations}
As discussed above, one limitation of our study is that it analyzed all CS standards
but not standards from other subjects.
Another limitation is the June 2024 cut-off date, due to collection, cleaning, and analysis being a very time-intensive process.
As standards do not update frequently, and no new states have released standards since then, we believe that the impact on the results is minimal.
%
Finally,
our study does not shed light on all the possible ways standards could be improved.
In particular, we leave for future work questions about the usability and understandability of standards, which are best answered by teachers. We further discuss the limitations of our work, especially as it pertains to our results, in Section~\ref{sec:discussion}.

\subsection{Positionality Statement}
Our team brings complementary expertise in K-12 CS teaching, security and privacy education research, and standards development: the first author has taught CS in U.S. K-12 settings, the second author has served on a standards committee, and the third author is an S\&P education researcher with 7+ years of experience. This provides insight into both classroom practice and the constraints of standards development, where security expertise is not always represented, and resources are limited.


\begin{table*}
\footnotesize
\centering
\begin{threeparttable}
\footnotesize
\setlength\extrarowheight{-6pt}
\setlength{\tabcolsep}{2pt}
\begin{tabular}{l@{ ~~} c c | l@{ ~~} c c}
    \textbf{Topic} & \multicolumn{1}{l}{\textbf{\# States}} & \multicolumn{1}{l}{\textbf{\# Standards}}  & \textbf{Topic} & \multicolumn{1}{l}{\textbf{\# States}} & \multicolumn{1}{l}{\textbf{\# Standards}}\\
    \toprule \toprule
\textit{\textbf{Encryption}} & 34 & 153 & \textit{\textbf{Organizational Security}} & 25 & 245\\
Broad coverage of encryption & 33 & 84 & Acceptable use policies/acceptable tech use &  21 & 105 \\
Encryption w/ public key & 11 & 20 & Methods for attack recovery & 10 & 37\\
Encryption w/ Caesar Cipher & 10 & 12 & Security at the organizational level & 8 & 20\\
Encryption w/ Steganography & 8 & 12  & Using \snp{} tool/process & 5 & 50\\
Encryption w/ other ciphers/types & 5& 15 &Digital forensics & 3 & 16 \\
Discussion of symm/asymm encryption & 5 & 8 & Threat modeling & 3 & 15\\
\cellcolor{myyellow!40}Encryption w/ RSA & \cellcolor{myyellow!40}2 & \cellcolor{myyellow!40}2 & \cellcolor{myyellow!40}Operating system security &\cellcolor{myyellow!40}2 & \cellcolor{myyellow!40}2\\
&&&&&\\
\textit{\textbf{Attacks \& Attackers}} & 37 & 312 & \textit{\textbf{Defenses}} & 41 & 534 \\
Malware & 26 & 54 & \cellcolor{mygreen!40}Physical (locks) and digital (encryption) defenses & \cellcolor{mygreen!40}37 & \cellcolor{mygreen!40}143\\
Broadly talk about attacks & 23 & 53 & Importance of passwords & 29&77 \\
Social engineering & 20 & 55 & Other methods of authentication & 26&59 \\
Viruses & 19 & 32 & Recommending general protections & 24 & 33  \\
Beneficial vs malicious hacking & 13 & 18 & Preventing unauthorized access and access control & 21 & 43 \\
Ransomware & 12 & 20 & Broadly discuss defense tradeoffs & 21 & 31 \\
Spyware & 10 & 14 & Evaluate usability vs security tradeoffs & 19 & 26 \\
Worms & 9 & 13 & Evaluate efficiency tradeoffs & 14 & 16 \\
Other attacks not covered & 8 & 25  & Evaluate defense tradeoffs & 13 & 16\\
Attackers’ motivations & 6 & 9 &  Evaluate feasibility tradeoffs & 13 & 15 \\
Types of attackers & 5 & 11  & CIA triad & 9&28 \\
Broadly talk about attackers & 3 & 6 & Recommend 2FA & 8 & 13 \\
\cellcolor{myyellow!40}Humans as a weak link & \cellcolor{myyellow!40}1 &\cellcolor{myyellow!40}2 &  Recommend biometrics & 7 & 11 \\
& & &Recommend token-based protection & 6 & 9 \\
& & & Defense in depth & 3&8 \\
&&&\cellcolor{myyellow!40}Recommend password requirements & \cellcolor{myyellow!40}1 & \cellcolor{myyellow!40}2 \\
&&&\cellcolor{myyellow!40}Recommend other protections & \cellcolor{myyellow!40}2 & \cellcolor{myyellow!40}2 \\
&&& \cellcolor{myyellow!40}Recommend geolocation & \cellcolor{myyellow!40}1 & \cellcolor{myyellow!40}2\\
&&&&&\\
\textit{\textbf{Networking Security}} & 35 &231 & \textit{\textbf{Safe \& Appropriate Behavior}} & 41&577\\
Networking security-related issues & 25 &105  & \cellcolor{mygreen!40}How to treat others online & \cellcolor{mygreen!40}37&\cellcolor{mygreen!40}225\\
Secure data transmission & 24&45& Keeping one's info private online & 25&69\\
Transmission w/ HTTPS & 15&23& Behaving securely online & 22&95 \\
Protection w/ firewalls & 9&26 & General S\&P practices for safety & 22&67\\
DDoS attacks & 8&12  & Keeping login info private & 20&45\\
Protection w/ other network defenses & 4 & 14 &Pros/cons of keeping info private & 18&26\\
\cellcolor{myyellow!40}Prevention of XSS & \cellcolor{myyellow!40}2&\cellcolor{myyellow!40}5 & Storing/handling data securely & 9&19\\
\cellcolor{myyellow!40}Internet availability& \cellcolor{myyellow!40}1&\cellcolor{myyellow!40}1 & Practices for safe network management & 8&10 \\
&&& Importance of updating software/apps & 7 & 13\\
 &&& Altering account settings & 7&8 \\
&&&&&\\
\multicolumn{1}{l}{\textit{\textbf{Laws and Intellectual Property}}} & 43&706 & \multicolumn{1}{l}{\textit{\textbf{Equity, Accessibility, and Diversity}}} &  44&818\\
\cellcolor{mygreen!40}Protecting intellectual property & \cellcolor{mygreen!40}41&\cellcolor{mygreen!40}272 & \cellcolor{mygreen!40}Consider ethics in tech & \cellcolor{mygreen!40}41&\cellcolor{mygreen!40}279\\
\cellcolor{mygreen!40}Laws and regulations & \cellcolor{mygreen!40}33&\cellcolor{mygreen!40}187 & \cellcolor{mygreen!40}Usability/accessibility &  \cellcolor{mygreen!40}38&\cellcolor{mygreen!40}176 \\
Credit and attribution & 33&104  & \cellcolor{mygreen!40}Impact of tech on equity &  \cellcolor{mygreen!40}33&\cellcolor{mygreen!40}74 \\
Credit for code re-use & 30&112 & \cellcolor{mygreen!40}Value of diversity &  \cellcolor{mygreen!40}31&\cellcolor{mygreen!40}137 \\
Tradeoffs w/ laws and regulations & 20&31 & Identifying bias in tech and oneself &  30&97 \\
& & &Reducing bias in tech & 24&31\\
& & & Altering tech to improve equity &  23&24 \\
&&&&&\\
\multicolumn{1}{l}{\textit{\textbf{Tracking and Digital Footprint}}} & 39&248 & \multicolumn{1}{l}{\textit{\textbf{Real World Impacts}}} &  37&152 \\
\cellcolor{mygreen!40}Broad discussion of tracking awareness & \cellcolor{mygreen!40}34&\cellcolor{mygreen!40}92 & Social impacts of privacy &  27&78 \\
\cellcolor{mygreen!40}Data permanence & \cellcolor{mygreen!40}31&\cellcolor{mygreen!40}145 & Current events in \snp{} & 21&37\\
Ways to prevent tracking & 4&11 & Social impacts of security & 20&32\\
&&& \cellcolor{myyellow!40}Soft skills needed in \snp{} & \cellcolor{myyellow!40}2 &\cellcolor{myyellow!40}5\\
&&&&&\\
\multicolumn{1}{l}{\textit{\textbf{Secure Coding}}} & 31 &146 & \multicolumn{1}{l}{\textit{\textbf{Content and Information}}} & 25&156 \\
Broad discussion of software security & 23&68 & Evaluating sources/data & 21&124 \\
Pros/cons of different programming languages& 19&22& Inappropriate content & 10&13 \\
Input validation & 11&18 & Internet censorship & 6&11\\
Bounds checking & 10&16 & Dis/misinformation & 5&8 \\
Secure coding practices & 7&17&&& \\
\cellcolor{myyellow!40}Circular references & \cellcolor{myyellow!40}4&\cellcolor{myyellow!40}4 & && \\
\cellcolor{myyellow!40}Key management & \cellcolor{myyellow!40}1&\cellcolor{myyellow!40}1 & &&\\
&&&&&\\
\multicolumn{1}{l}{\textit{\textbf{Troubleshooting}}} & 39&304 & \multicolumn{1}{l}{\textit{\textbf{Code Debugging}}} & 36&298 \\
\cellcolor{mygreen!40}Basic troubleshooting (know something is wrong) & \cellcolor{mygreen!40}38&\cellcolor{mygreen!40}236 & \cellcolor{mygreen!40}Testing code& \cellcolor{mygreen!40}36&\cellcolor{mygreen!40}141\\
Advanced troubleshooting (know what is wrong) & 29&68 & Debugging code & 29&139\\
& && Performing code review&  15&18 \\
&&&&&\\
\multicolumn{1}{l}{\textit{\textbf{Cloud Computing}}}
& 13&29 & \multicolumn{1}{l}{\textit{\textbf{Embedded Systems}}} & 23&71 \\
Cloud computing & 12&29 & Embedded systems &  23&71\\
Cloud computing security & 3&9 & Embedded systems security & 5&6 \\
&&&&&\\
\multicolumn{1}{l}{\textit{\textbf{Networking}}} & 38 &566 & \multicolumn{1}{l}{\textit{\textbf{Other}}} & 22 & 95 \\
\cellcolor{mygreen!40}How data travels & \cellcolor{mygreen!40}36&\cellcolor{mygreen!40}197 & Security but no specifics & 21&75 \\
\cellcolor{mygreen!40}Broad discussion of networking basics & \cellcolor{mygreen!40}34&\cellcolor{mygreen!40}219 & Privacy but no specifics & 5&20\\
Non-security networking issues & 29&70 &  &&\\
Network structures & 27&80  & & &\\
\bottomrule \bottomrule
\end{tabular}
\end{threeparttable}
\caption{Topics and the number of states/standards covering them. Most and least common topics are highlighted in \textcolor{mygreen}{green}/\textcolor{myyellow}{yellow}, respectively.}
\label{tab:topics}
\end{table*}

\section{Analysis of CS teaching standards}
In this section, we detail the \snp{} topics covered by the standards (\textbf{RQ1}), the organization of the standards (\textbf{RQ2}), and variance in the standards across states and grade levels (\textbf{RQ3}).

Of the 47 states that had computer science standards,\footnote{%
States that had no computer science standards were Vermont, Maine, and Oregon. Oregon has released a draft of computer science standards for public comment~\cite{OregonDeaprtmentOfEducation}.} 
37 created their own or adapted existing standards, six published the CSTA standards on their website with some form of state branding, three adopted the CSTA standards and provided a link, and one did the same for both the CSTA and ISTE standards.
Between the unique or adapted state standards and the standards created by standards bodies, we analyzed 45 total unique sets of standards (37 unique/adapted state standards and eight national organization standards). In total, we analyzed \totalStandards{} standards, \securityStandards{} of which we classified as security- and privacy-related (\percentSecurity{}\%).

We identified \elementarySecurityStandards{} security- and privacy-relevant standards at the elementary level, \middleSecurityStandards{} at the middle school level, and \highSecurityStandards{}  at the high school level. We identified 241 security- and privacy-relevant standards that did not have a marked grade level\footnote{%
The sum of these is greater than \securityStandards{} because some states did not use these grade bands. If a standard was marked 6--12, it was counted for both middle school and high school.} 
or were from ISTE, which also specified standards for teachers and administrators.\footnote{%
These were analyzed alongside the standards for students, since Connecticut adopted both ISTE and CSTA.} On average, each state had 271 total standards (221 if Georgia,%
\footnote{Georgia specifies standards for more individual courses than any other state, including a dedicated Cybersecurity course.}
which had 2,453 total standards, is excluded).
Only one set of standards (IB Computer Science Option D) did not contain any security- or privacy-related content.  

\subsection{\snp{} topics covered by CS standards (\textbf{RQ1})}
Standards covered \numCodes{} security- and privacy-relevant topics, most of which were technical, like encryption, and some of which were social, like equity, accessibility, and diversity. 
All topics and the number of states and standards that covered them can be found in Table~\ref{tab:topics}. 


\summary{States teach dangers of tracking---and benefits.}
Unsurprisingly, almost all states\footnote{%
    For brevity, we refer to states and organizations collectively as ``states.''
} had standards about keeping students and their data safe. This was often framed in terms of practicing helpful behaviors and preventing harmful behaviors, though most standards did not specify which behaviors to teach. Some specified areas of focus, like online safety and settings management. 

Around two-thirds of states had standards instructing students to be careful with their online reputations and that information shared online can be permanent. This was sometimes framed as a way of protecting privacy: \standard{Explain the connection between the longevity of data on the internet, personal online identity and personal privacy}{Maryland}. Elsewhere, it was framed as a way of protecting one's reputation: \standard{Students will demonstrate an understanding of the importance of creating and maintaining a positive online identity and the permanence and future impact of their online and offline decisions when using digital technology}{South Dakota}. 

More than 75\% of states had standards covering the presence and scale of online tracking. Some framed this as a privacy concern: \standard{Explain the privacy concerns related to the collection and generation of data through implicit and explicit processes}{Indiana}. Others presented it as a process for students to understand: \standard{Describe ways web advertising collects personal information}{Alabama}.
Only four states had standards that covered tracking prevention. 

While more than half of states had standards stating that some information should be kept private, almost 40\% of states had standards indicating that there were benefits to sharing information online, such as: \standard{Discuss trade-offs such as privacy, safety, and convenience associated with the collection and large scale analysis of information about individuals (e.g., social media, online shopping, how grocery/dept stores collect and use personal data)}{Idaho}.
The Advanced Placement curriculum appeared to go even further: \standard{PII and other information placed online can be used to enhance a user’s online experiences}{APCSP}.

\summary{Almost all states cover ethics and accessibility.}
We coded societal topics such as diversity, ethics, and accessibility as security-relevant and found that they were common among CS standards.
Ethics was covered in more than 91\% of states; around 84\% covered usability and accessibility, 73\% covered equitable access to technology (e.g., \standardP{Evaluate the impact of equity, access, and influence on the distribution of computing resources in a global society}{Wisconsin}), and around 70\% covered bias and diversity. 
More than half of states had standards that asked students to take a computational artifact and make it less biased or more equitable. Standards mentioned ethics in many contexts, from defenses (see below), to laws and regulations (e.g., \standardP{Evaluate the evolving legal and ethical tradeoffs that shape computing practices}{South Carolina}), to student behavior when using technology (e.g., \standardP{Students will practice positive, safe, legal, and ethical behavior when using technology}{South Dakota}).

\summary{States emphasize digital citizenship and personal online safety.}
More than 91\% of states had standards covering safe and appropriate online behavior. Most focused on digital citizenship and how students should interact with others online, often emphasizing respectful, ethical, or responsible technology use: \standard{Practice responsible digital citizenship (legal and ethical behaviors) in the use of technology systems and software}{Idaho}. 

Around 55\% of states specifically discussed protecting personal information online, including the importance of remaining anonymous in some contexts and avoiding the sharing of personally identifiable information: \standard{Manage personal data to maintain digital privacy and security and be aware of data--collection technology used to track online behaviors}{North Dakota}. About half of states additionally included standards related to practicing secure online behaviors more broadly, typically focusing on general privacy and security habits students could use to protect themselves online (e.g., \standardP{With support and guidance, identify the importance of staying safe while using family-and educator-approved Internet sites}{West Virginia}). However, many of these standards remained high-level and did not specify the concrete behaviors, tools, or practices students were expected to learn.

\summary{States frequently discuss intellectual property, but less often discuss legal trade-offs.}
Laws, regulations, and intellectual property were covered in more than 95\% of state standards. 
Most commonly, standards focused on respecting and protecting intellectual property, often through discussions of attribution, copyright, licensing, or appropriate technology use: 
\standard{Describe how different types of software licenses (e.g., open source and proprietary licenses) can be used to share and protect intellectual property}{Florida}.

Some states additionally referenced laws and regulations in a broader sense, generally encouraging discussion of legal or ethical considerations surrounding technology use,
(e.g., \standardP{Students will practice positive, safe, legal, and ethical behavior when using technology}{South Dakota}).
However, only 44\% of states included more concrete discussions of laws, regulations, or policy trade-offs. These standards were more likely to encourage students to debate the pros and cons of laws and regulations, and the specific positive and negative impacts of intellectual property laws:
\standard{Explain the beneficial and harmful effects that intellectual property laws can have on innovation}{Washington}.

\summary{Encryption is painted in broad strokes.}
Roughly three quarters of states had standards about encryption, though most detailed it only in broad terms:
\standard{Implement an encryption, digital signature, or authentication method}{Florida}.
Specific examples tended to be historical (e.g., Caesar, Bacon's, or Vigenère cipher).
While eleven states had standards that mentioned public-key encryption, only two named a specific cryptosystem (RSA).
In comparison, eight states had standards that focused on steganography.

\summary{Standards cover a breadth of defenses and attacks.}
States' standards indicated that students should be able to choose defenses against attacks.
Slightly over half of states want students to be able to recommend protections, though often without mentioning a specific scenario
(e.g., \standardP{Recommend security measures to address various scenarios based on factors such as efficiency, feasibility, and ethical impacts}{Illinois}).
A little over 46\% of states further emphasized the need to consider trade-offs as part of the recommendation process
(\standardP{Explain tradeoffs when selecting and implementing cybersecurity recommendations}{CSTA}). A few states had standards that mentioned specific defenses, most of which related to authentication (e.g., 2FA, biometrics), though these were mostly a list of examples in a standard that was primarily about recommending or evaluating defenses. When evaluating defenses, state standards had text encouraging students to consider feasibility, efficiency, and usability. Of these, usability was the most popular, occurring in around 42\% of states. 

Standards also covered social engineering, malware, ransomware, viruses, spyware, worms, and people as an attack vector. 
DDoS also appeared in these lists, but other networking security topics did not. Malware (around 58\% of states), social engineering (around 44\% of states), and viruses (around 42\% of states) were the most common attacks discussed, but it was equally common to discuss attacks without specifying specific ones to teach (around 51\% of states): \standard{Research and describe common attacks on hardware, software, and networks}{Arkansas}. Attackers were rarely discussed, most often in the context of white-hat versus black-hat hacking. 

\subsection{Organization of \snp{} standards (RQ2)}
In this section, we detail our analysis of the presence of examples, and how standards were organized.

\summary{Examples ranged from missing to detailed.}
As discussed in Section~\ref{ss:snp_k12},
we noted whether standards had examples or clarifying statements,
since these provide essential guidance to teachers,
especially those who may lack prior \snp{} background.
Of the \securityStandards{} \snp{} standards we identified, \hasExamplesSecurity{}---from \hasExamplesStates{} different states---contained examples.
For instance, the standard 
\standard{Distinguish between private vs. public information}{CISA} had an accompanying ``\textit{Clarification statement: At this level, student discussions should focus on grade-appropriate examples of privacy and what is OK to share about themselves and how that relates to confidentiality.}'' 

The format of examples varied, although they tended to be consistent within states. Examples fell into three major patterns.
Seven states included clarifying statements to explain what should be taught, though they did not always label them as such or include them for all standards. 
%
Five states provided specific examples of how a topic should be taught;
e.g., one from California specified an activity: ``\textit{Students could devise a plan for sending data to represent a picture, and devise a plan for interpreting the image when pieces of the data are missing.}''
California was also the only state to include definitions of terms and a statement of intent. 

The most common type of example (18 states) was a list of terms that fell under a given topic. For example, \standardP{Explain principles of network security and techniques that protect stored and transmitted data (e.g., encryption, cryptography, and authentication)}{West Virginia}. Some states only had one or two standards which contained the list of terms.

\summary{\snp{} standards were not always labeled as such.}
While most state standards were grouped under categories (e.g., Networks \& the Internet, Algorithms \& Programming, Impacts of Computing), only \numSecStates{}~state standards had categories explicitly labeled as security and/or privacy, despite 44 states having standards that covered this topic. Of the \securityStandards{} \snp{}-related standards we identified, only \securityCategoryStandards{} were categorized by their state as related to \snp{} (\securityCategoryStandardsPercent{}\% of all standards, \percentOfSecurityCalledSecurity\% of all standards we labeled as \snp{}). 
For example, we categorized tracking as \sap{}, but states generally categorized it as ``Impacts of Computing'' or ``Digital Citizenship.'' When states did not have a security category, security standards often fell under ``Networks \& the Internet.'' States generally labeled encryption, attacks \& attackers, defenses, and organizational security as security.

\summary{Secure coding was sometimes mixed in with other topics.} 
States were divided on whether or not to label software security as security or place it with other programming standards. For example, Alaska labeled \textit{Compare ways software developers protect devices and information from unauthorized access} as Cybersecurity and \textit{Explain security issues that might lead to compromised computer programs} as Program Development. Generally, specific actions that programmers should take (such as secure coding, bounds checking, and input validation) were not categorized as security and instead fell under program development. Software security (e.g., \standardP{Utilize a software lifecycle process that considers security to plan and develop programs for all types of users}{Maryland}) and key management were the only related topics that were categorized as \snp{} in more than half of occurrences, though key management only occurs once.

\summary{States labeled security more often than privacy.} 
Topics referencing privacy were often not categorized as such. Only four states had any standards that were categorized as privacy. 
Privacy-relevant topics such as keeping information private, the importance of curating your online presence, and the societal implications of privacy were only categorized as \snp{} 50\%, 24.7\%, and 16.1\% of the time, respectively. 
This may make it difficult for teachers creating a unit about privacy to easily find all relevant standards.

\subsection{Variance of \snp{} topics (RQ3)}
In this section, we detail how the topics covered in the standards varied between states and grade levels. 

\subsubsection{Variance between states}
\begin{figure*}
        \includegraphics[width=1\linewidth]{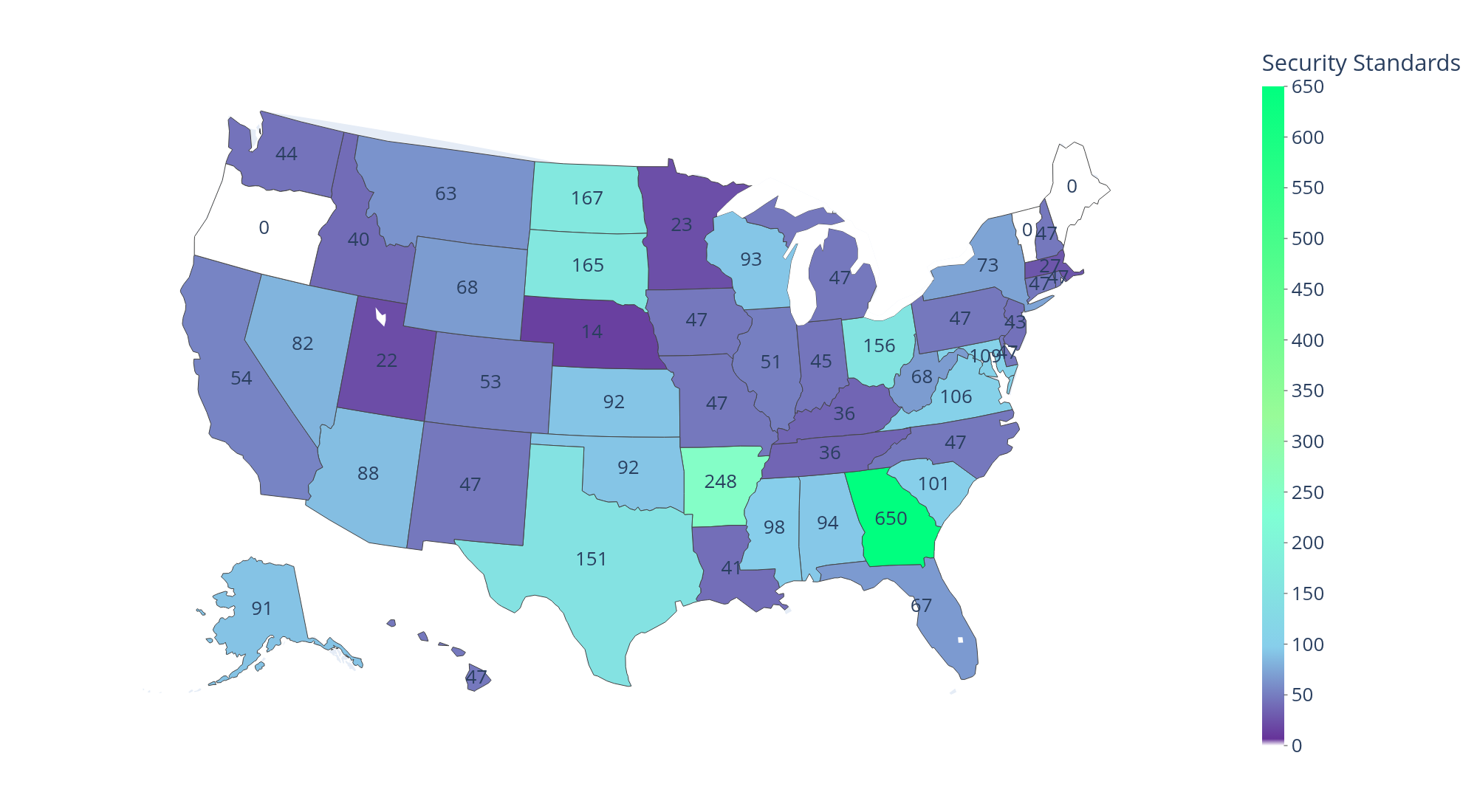}
        \caption{%
            Number of \snp{} standards per state. States in white had no CS standards.
        }
        \label{fig:standards}
        
    \end{figure*}

Topic coverage between states varied widely
(Figure~\ref{fig:standards}).
We identified
the 10 topics with the most and fewest standards and the 10 that occurred in the most and fewest states,
yielding (due to overlap) a total of 14 most common and 11 least common topics. 
The most commonly covered topics include testing code, ethics, and credit and intellectual property.
They are highlighted in green in Table~\ref{tab:topics}. The least commonly covered \codes{} are highlighted in yellow in Table~\ref{tab:topics} and include RSA encryption, web availability, and key management. 

\summary{Even the most popular topics were not universal.}
Thirty-two states had standards that were missing one or more of the most common topics.  
Interestingly, every single set of standards from non-state entities was missing at least one of the most common topics,
and APCSA (11), IB Standard Level (9), and ISTE (6) had the most missing.
The only state missing more in their standards was Minnesota, with eleven of the fourteen most common topics absent.
Ten states had one or more of the least common \codes{} covered by their standards, with Texas (7), CISA (2), Colorado (2), and Alabama (2) having the most.

\label{sec:results-audience}
\summary{Most topics are geared towards general users.}
As described in Section~\ref{sec:methods-analysis}, we categorized topics as being
relevant to general technology users (e.g., passwords, safe behavior online)
and those most relevant to future technology professionals (e.g., digital forensics, input validation).
Except for six states, most included more standards that benefit all students rather than only those pursuing technical careers.
All of the least commonly taught \codes{} were aimed at future technology professionals, which may be why Texas had so many: only Georgia and Texas specifically outline a course in cybersecurity as part of their CS standards.
%

\begin{figure*}
        \centering
        \hspace*{-0.10in}
        \includegraphics[width=1\linewidth]{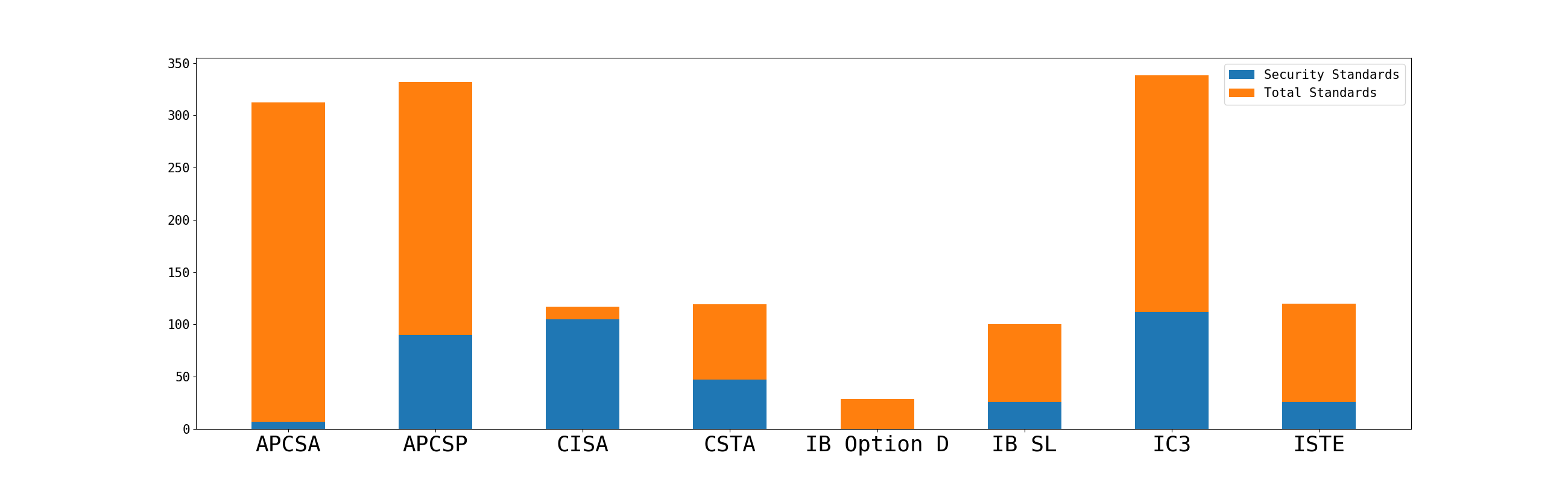}
        \caption{%
            Number of CS and \snp{} standards per organization.
        }
        \label{fig:orgsonly}
        
    \end{figure*}

\summary{National-level standards vary heavily in topic coverage.}
(See Figure~\ref{fig:orgsonly}.)
CISA and CSTA standards covered more \codes{} than most states
(60 and 56, respectively, out of \numCodes{} total topics),
with only Texas (92), Georgia (86), and Kansas (71) covering more.
The other national standards covered fewer topics, with IC3 and APCSP covering 30 and 34 topics respectively and APCSA, IB Option D, and IB Standard Level all covering fewer than 10. 

\subsubsection{Variance between grade levels}
The subjects that standards focused on varied with grade level.

\summary{Fundamentals are prioritized in elementary schools.}
Acceptable use and general safety were most commonly found in elementary level standards. For example, passwords and logins were most often taught at the elementary level, and dropped off sharply as students grew older. Acceptable use and appropriate behavior online were also emphasized at the elementary school level, but the drop-off in later grades was less sharp: high school students were still expected to \standard{demonstrate and advocate for positive, safe, legal, and ethical habits when using technology and when interacting with others online}{West Virginia}. 

\summary{Secure transmission and networking are prioritized in middle schools.}
Web security (e.g., \standardP{Identify existing cybersecurity concerns with the Internet and systems it uses}{Alaska}) and 
secure data transmission (e.g., 
\standardP{Apply multiple methods of encryption to model the secure transmission of information}{Mississippi})
were covered most often in middle school standards. 
Foundational versions of technical topics that were covered at the high school level were also sometimes mentioned here. For example, in some states, standards covering how to identify defenses (e.g., \standardP{Explain physical and digital security measures that protect electronic information}{Rhode Island}) were present in middle school standards. Then, at the high school level, students were expected to recommend defenses for different scenarios and explain their reasoning (e.g., \standardP{Explain tradeoffs when selecting and implementing cybersecurity recommendations}{Rhode Island}). 

\summary{Technical topics and critical thinking were most common in high school standards.}
Secure coding, embedded systems, and real-world security and privacy were covered most frequently at the high school level. Attacks and attackers was also present most often in high school standards, with the exception of social engineering, which was covered more often in middle school standards. 
Debate and discussion were considered most suitable for high school students. For example, laws and regulations were covered in at least 15 states at every grade level, but the trade-offs associated with regulation (e.g., \standardP{Explain the beneficial and harmful effects that intellectual property laws can have on innovation}{Illinois}) were covered in 18 states at the high school level, two states at the middle school level, and one state at the elementary level.

\summary{Social and societal topics were covered at all levels.}
On the social side, intellectual property \& credit and usability \& accessibility were covered at all grade levels. Other topics were covered most heavily at one grade level and then covered again, less frequently, in other grade levels. Diversity and inclusion was present most often at the elementary-school level. Ethics was covered at all levels, but most frequently in high school. In middle school, standards covered identifying bias (e.g., \standardP{	
Discuss issues of bias and accessibility in the design of existing technologies}{Montana}), and in high school, they covered reducing bias and understanding and improving equity (e.g., \standardP{Test and refine computational artifacts to reduce bias and equity deficits}{Montana}).

\section{Discussion}
\label{sec:discussion}
\label{sec:discussion}
As an initial exploration of \snp{} teaching standards in the United States, our results shed light on what topics states would like to be taught to K-12 students.

\subsection{From Standards to Classroom Implementation}
Standards documents primarily define \emph{what} students should learn rather than \emph{how} concepts should be taught and therefore provide only a partial view of students' actual educational experiences. While many U.S. school districts adopt relatively standardized instructional materials for core subjects, this practice is substantially less consistent in computer science education. The issue may be particularly pronounced for security and privacy topics, where teachers often develop, adapt, or independently source instructional materials due to the limited availability of widely adopted curricula and the rapidly evolving nature of the field.

As a result, there may be substantial variation between intended curricula represented in standards and enacted curricula implemented in classrooms. Differences in teacher expertise, instructional autonomy, district-level priorities, resource availability, student populations, and enforcement mechanisms may all influence how \snp{} concepts are ultimately taught. Consequently, the presence of a topic within standards does not necessarily imply consistent instructional coverage, depth, or pedagogical quality across schools and districts.

Accordingly, our work focuses specifically on intended curricular content rather than classroom implementation. However, understanding how \snp{} standards are translated into practice represents an important direction for future research. Future work should examine enacted curricula through analyses of instructional materials, classroom observations, teacher interviews, assessments, and professional development practices to better understand how students actually encounter \snp{} concepts in K-12 education.

\subsection{State Standards Overlap, but No Topics Are Universal}
State \snp{} standards covered a large range of topics---\numCodes{}, according to our categorization.
At a high level, there was a considerable degree of agreement about the subject areas that should be taught.
For example, 37 out of 45 states had standards that discussed different types of attacks and attackers,
and 41 out of 45 had standards covering various defenses.
Yet, it is notable that no topics were covered by all states,
and even the most popular topics were covered by only two thirds of states and organizations.
This leads us to conclude that there is no complete consensus about what should be taught in security and privacy.
On one hand, this is not surprising:
some variation between states is always expected,
and prioritization of \snp{} topics is notoriously difficult even for experts~\cite{Redmiles20}.
On the other hand, security threats are the same regardless of the state where one grows up,
and the fact that different students are receiving diverging instruction suggests that there may also be a discrepancy in outcomes.
Future research should examine student outcomes by state
and investigate the possibility of more empirically driven standards.

\subsection{Balancing Enduring Concepts with Rapidly Evolving Technologies}
Not all security and privacy topics age at the same rate. Some foundational concepts, such as data protection, digital footprint, and responsible online behavior represent enduring principles that are unlikely to change substantially even as technologies evolve. These topics may therefore be well suited for inclusion directly within long-term standards documents.

In contrast, other topics identified in the standards may have substantially shorter instructional lifespans because they are closely tied to specific technologies, platforms, tools, or contemporary digital practices. Examples involving social media platforms, privacy settings, device ecosystems, cybersecurity tools, attack techniques, or platform-specific security features may become outdated quickly as technologies and user behaviors evolve. Similarly, instructional examples centered around current online trends or specific applications may lose relevance within only a few years. Embedding highly transient material directly into statewide or national standards may therefore reduce the longevity and maintainability of standards documents.

Artificial intelligence may further accelerate this challenge. Many emerging \sap{} concerns related to AI connect directly to topics identified in our standards analysis, including misinformation, privacy, data collection, and online safety. Rapid advances in generative AI systems, AI-assisted software development, automated phishing and social engineering, deepfakes, and AI-driven surveillance are already reshaping both security threats and privacy risks~\cite{lee_deepfakes_2024}. At the same time, AI technologies themselves are evolving rapidly, making it difficult for static standards to remain current. Topics that may be highly relevant today---such as prompt injection, synthetic media detection, AI-generated misinformation, or privacy risks associated with large language models---may change significantly as AI systems, regulations, and public usage patterns evolve.

These dynamics suggest that standards bodies should distinguish between enduring conceptual principles and rapidly changing technological examples. Stable concepts may be appropriate for inclusion directly within standards, while shorter-lived technologies, tools, and contemporary case studies may be better addressed through supplementary instructional materials that can be updated more frequently. This approach may allow standards to retain long-term stability while still enabling educators to address emerging \snp{} risks and technologies in a timely manner.

\subsection{Vagueness Is Strategic but May Create Difficulties for Teachers}
One distinctive aspect of standards that we observed, though did not formally analyze, is their vagueness.
The strategic ambiguity of a standard like
\standard{Compare ways to protect devices, software, and data}{Alabama}
means that a teacher can still illustrate this topic with different defenses as technology evolves.
However, doing so also increases demands on the teacher:
they must know enough about the subject to be able to turn the vague guideline into a concrete lesson plan,
then keep it up-to-date with changing technology.
Standards could help by including more examples---something our analysis found was lacking. However, introducing examples may reduce the longevity of the standard. 
States and national organizations could also provide sample curricula and modules.
Another solution is to recruit more teachers with relevant \snp{} background and to provide training for those who do not already have it.

\subsection{Supplementary Materials May Reduce Ambiguity While Supporting Adaptability}

A promising approach for balancing flexibility with instructional support is the use of supplementary materials accompanying standards documents. In practice, some states and organizations already provide supporting resources beyond the standards themselves, including curriculum frameworks, implementation guidance, lesson examples, and instructional materials. However, the availability, specificity, and quality of these resources appear to vary substantially across states and organizations. Our findings suggest that such supplementary materials may play an important role in helping teachers interpret and implement broadly worded \snp{} standards.

Rather than embedding extensive examples or technology-specific guidance directly within standards, states and standards bodies could provide separate supporting resources that include illustrative lesson plans, classroom activities, implementation examples, glossaries, mappings to learning objectives, and example technologies or case studies. This approach preserves flexibility within the standards themselves while still providing educators with concrete guidance for classroom implementation.

Supplementary materials may help address several challenges identified in our analysis. First, they may reduce ambiguity by providing teachers with clearer interpretations of broadly worded standards and more concrete examples of how \snp{} concepts might be taught in practice. Second, maintaining supporting materials separately from standards may lower the barrier for updates as technologies, platforms, threats, and best practices evolve. Updating supplementary resources is often substantially easier and faster than revising statewide or national standards documents, which may operate on multi-year revision cycles.

This distinction may be particularly important for \snp{} education because some topics evolve much more rapidly than others. Foundational concepts such as authentication, personal data protection, or secure communication may remain relatively stable over time, whereas discussions involving specific social media platforms, authentication technologies, attack techniques, privacy tools, AI technologies, or emerging digital platforms may become outdated quickly. Supplementary materials may therefore provide a practical mechanism for preserving stable conceptual standards while still enabling instructional guidance to adapt to rapidly changing technological contexts.

\subsection{Teachers Are Central}
Our analysis illustrates the crucial role of teachers,
who are tasked with interpreting the often-vague standards and turning them into detailed curricula and lesson plans.
Are the standards too vague or just right?
It all depends on the teacher.
We also acknowledge teachers as a missing element of this study.
We see their perspective as critical,
but we felt that incorporating an additional set of interviews with the existing components of this work would dilute the results of each.
We therefore leave this for future work,
noting, however, that our team includes a researcher with significant high school teaching experience,
which has shaped our approach at all stages of research.

\subsection{Standards Need a Clearer Audience and Goals}
Our analysis highlighted a persistent point of confusion with the standards:
who are they for?
Students in high school CS classes might become security professionals, work in software more generally, or end up as neither of those.
Which of these paths should the course be preparing them for?
A similar dilemma faces those developing standards for CS in general, as well as other topics.
However, this question is especially acute in \snp{}.
Topics like secure coding are important for future developers and appropriate even in an introductory CS class.
However, staying safe online is even more fundamental.
Given limited classroom time, having a clear vision for the educational goals of the class (and associated standards) can help with prioritization.
This is something current standards seem to be lacking.
Future standards would do well to more clearly articulate their goals.
One approach might be to divide standards into mandatory courses and elective courses, so that future tech professionals can learn the \sap{} topics they need while all students get a necessary background in \sap{}.


\subsection{General Challenges Versus \snp{}-Specific Challenges}
Many of the observations identified in this work---such as vague standards language, limited examples, inconsistent specificity, and uneven implementation guidance---are not unique to security and privacy education. Similar challenges have been identified more broadly in K-12 computer science education, particularly when translating intended curricula into enacted classroom practice~\cite{Falkner19}. More generally, these issues are also common across curriculum design efforts in emerging or interdisciplinary subject areas where standards evolve more rapidly than instructional infrastructure.

However, our findings suggest that these challenges may be uniquely intensified in the context of \snp{} education. Unlike many traditional K-12 subjects, U.S.\ computer science teachers often lack formal preparation in \snp{} topics specifically. Prior work has shown that teacher preparedness, available resources, and interpretation of curricular expectations substantially influence computing curriculum implementation~\cite{Samarasekara22}. In \snp{}, these concerns may be amplified because the field evolves rapidly and requires teachers to interpret technically complex and socially contextualized topics related to security, privacy, online safety, digital risk, and responsible technology use.

As a result, ambiguities that may be manageable in other areas of computer science education can create disproportionately large instructional challenges for \snp{} topics. For example, vague references to privacy or security without accompanying examples, learning objectives, or implementation guidance may leave teachers uncertain about the expected technical depth, scope, or pedagogical framing of the material. Because many \snp{} topics also intersect with law, ethics, policy, and digital citizenship, teachers may need to make additional interpretive decisions that extend beyond traditional computing instruction. This may contribute to inconsistent classroom coverage across schools and districts, particularly in contexts where teachers have limited prior exposure to \snp{} concepts.

Importantly, some implications of our findings are specific to \snp{}, while others apply more broadly to computer science education or curriculum design in general. The need for clearer examples, implementation guidance, and more consistent standards language reflects broader curriculum-design challenges that extend beyond \snp{} and even beyond computer science. In contrast, the implications related to rapidly evolving technical content, interdisciplinary interpretation, and the specialized expertise required to teach security and privacy concepts are more uniquely associated with \snp{} education. Our recommendations should therefore be interpreted across these different levels: some address general curriculum quality, some target broader K-12 CS implementation concerns, and others specifically address challenges unique or exacerbated within \snp{} instruction.

\subsection{Implications for International Security and Privacy Education}
Although our work focuses on the United States, the results provide several broader lessons for countries seeking to evaluate or expand security and privacy subjects within primary and secondary education. First, our analysis demonstrates the importance of examining \sap{} content systematically and explicitly rather than assuming these topics are implicitly covered through general computing instruction. Our findings showed substantial variation in both the presence and emphasis of \snp{} topics across states, suggesting that, even within a single country's educational system, there is no universal agreement about what students should learn. Other countries may therefore benefit from evaluating not only whether \snp{} topics appear in standards, but also how concretely they are specified and how consistently they are implemented.

The U.S. case study also highlights the broader challenge of curricular variation across educational systems. Prior international work has similarly found substantial differences in how computing education is structured and enacted across countries~\cite{Falkner19}. These differences suggest that international analyses of \snp{} education should carefully account for governance structures, regional variation, and local implementation practices rather than treating national curricula as uniform.

It is also likely the case that \snp{} concepts are not confined to computer science curricula alone. Instead, topics such as online safety, privacy awareness, misinformation, digital citizenship, and responsible technology use may be incorporated into media literacy, civics, information technology, or broader digital competence education.
As a result, analyses that focus exclusively on computer science standards may underestimate the extent to which students are exposed to \snp{} concepts in practice.
Rather than examining computer science standards in isolation, researchers should also consider adjacent subjects where security- and privacy-related competencies may be taught.

Additionally, not all countries may have a model like the U.S. of educational authorities publishing learning expectations in the form of teaching standards.
They may rely, for example, on common curricula or other instruments intended to ensure consistency.
Nonetheless, these may suffer from similar challenges as standards.
International research examining enacted computing curricula suggests that local factors, such as classroom implementation and teacher preparedness, may substantially shape what students ultimately learn~\cite{Falkner19, Samarasekara22}.
Accordingly, future work should adopt a broader curricular perspective when conducting cross-national analyses of \snp{} education. Such work would enable more accurate international comparisons and help determine whether students receive coherent and comprehensive \snp{} instruction across the curriculum or only fragmented exposure distributed across multiple subjects.

\subsection{\snp{} Are for Everyone, but K-12 CS Classes Are Not}
Security and privacy education is important for everyone,
but the \sap{} topics covered by the standards we analyzed are not reaching the majority of students.
Code.org estimates that only 6.1\% of high school students were enrolled in a CS course in the 2024-2025 school year,
with only 60\% of American high schools offering elective CS courses---and even fewer in low-income districts~\cite{StateComputerScienceEducation2024}. 
Since most of the U.S.\ population will not go to college, K-12 remains the best place to inculcate critical security and privacy skills. 

At the same time, relying exclusively on standalone CS courses to deliver \sap{} education may inherently limit its reach. Many \snp{} concepts naturally intersect with other core subjects and may therefore be suitable for broader curricular integration. For example, topics such as misinformation, digital footprints, online safety, responsible information sharing, and privacy awareness align closely with media literacy, digital citizenship, civics, and health education. Discussions surrounding data collection, algorithmic decision making, surveillance, and the societal impacts of technology may fit naturally within social studies or interdisciplinary technology courses. Integrating selected \snp{} topics across multiple subjects may help ensure that all students encounter foundational security and privacy concepts, including those who never enroll in a dedicated CS course.

However, expanding \snp{} education will require substantial resources and institutional support. School schedules are already crowded, teachers are frequently overloaded, and many educators may lack formal preparation in \snp{} topics. As a result, meaningful integration of \snp{} concepts will likely require not only curricular changes, but also professional development opportunities, instructional resources, and clearly articulated rationales for why these topics should be prioritized alongside other educational goals.

\section{Conclusion}
\label{sec:conclusion}
We analyzed 12,201 computer science teaching standards, identifying 3,920 that covered 109 \sap{} topics. Topics ranged from encryption to appropriate online behavior. The standards we analyzed often lacked concrete examples of what to teach, and the topics covered vary heavily by state and grade level. 
We recommend that standards are designed for a specific target audience: either the general population of students or students with a strong interest in technology. We also recommend that standards include examples, which should be updated more frequently than the standards themselves.

\section*{Acknowledgments}
We thank the anonymous reviewers who provided helpful
feedback on this paper and Abby Bissell-Westlake who assisted in the analysis of standards. 

\cleardoublepage
\appendix
\section*{Ethical Considerations} 
We structure the ethical considerations discussion as a stakeholder analysis, where we identify stakeholders, evaluate risks, and explain mitigation strategies. 

\subsubsection*{Stakeholder Analysis} Direct stakeholders of this research include states and standards committees that adopt our recommendations. Indirect stakeholders include teachers, students, and parents whose states or districts adopt our recommendations.


Our work involves collecting and reviewing standards published online by states and organizations. We did not obtain consent prior to reviewing standards, but as they were intended for public viewing and access, we did not see this as necessary to proceed with the research. Our work may add burdens to states and standards committees, as we recommend changes to standards structure and indicate weaknesses in published material. We believe, however, that those burdens will be outweighed by the benefits of improved standards to students and teachers, as well as the benefit of future work in this space providing peer-reviewed academic guidance on adding security and privacy to computer science standards.

\subsubsection*{Ethical Principles}

\summary{Beneficence.} We understand that additional demands on teachers and students may be harmful, especially in the context of the U.S.\ education system, where students and teachers are already overburdened. We seek to make recommendations that work within existing frameworks, that will reduce burden on teachers (e.g., recommending that states provide sample curricula and lesson plans to teachers, or provide training within existing professional development frameworks) and not increase student workload. We believe that studying and improving security and privacy education will benefit students, teachers, and society as a whole. 

\summary{Respect for Persons.} 
This work examines publicly available K--12 computer science standards and does not involve human subjects or the collection of personally identifiable information. Nevertheless, the analysis is motivated by respect for students, teachers, and educational stakeholders who may be affected by how security and privacy concepts are represented within educational systems. In particular, we recognize that teachers often operate under substantial curricular and institutional constraints, and our findings should not be interpreted as criticisms of individual educators or schools. Instead, the goal of this work is to better understand systemic patterns in \snp{} standards and identify opportunities to support more equitable, accessible, and effective \snp{} instruction. We also acknowledge that curricular decisions involve balancing competing educational priorities, resource limitations, and local needs, and therefore avoid prescribing a single universal approach to \snp{} education.

\summary{Justice.}
This work is motivated in part by concerns about equitable access to security and privacy education. Because computer science courses are not universally available in U.S. K--12 education, and because participation in CS courses varies substantially across schools and student populations, many students may never encounter formal \snp{} instruction. These disparities may be particularly pronounced in under-resourced or low-income districts where CS offerings are less common. Our analysis therefore considers how reliance on elective CS pathways may unintentionally limit access to important \snp{} concepts and skills. At the same time, we recognize that schools and educators operate under unequal resource constraints, including differences in funding, teacher preparation, curricular flexibility, and access to instructional materials. Accordingly, our recommendations aim to support broader and more equitable access to \snp{} education without assuming uniform institutional capacity across educational settings.

\summary{Respect for Law and Public Interest.} 
By understanding and advancing the creation of computer science teaching standards, this work serves the public interest by encouraging security and privacy education for students, preparing them for the increasing security and privacy challenges of the modern world. 

\summary{Publication Impacts.} Our research may influence future research directions, increasing interest in K-12 security and privacy education. We believe this will be more beneficial than harmful, resulting in improved understanding of what laymen know about security and privacy, as well as improved guidance for states and standards committees, and improved education for students. Our recommendations may influence states or teachers with an interest in security and privacy education. We kept this in mind when designing recommendations, endeavoring to place the burden of implementation on states and districts rather than individual teachers or students. We acknowledge that teachers are already overburdened and endeavor not to make recommendations that will contribute to this problem. 


\subsubsection*{Decision to Conduct and Publish} This research was initiated to better understand how security and privacy education is conducted at the K-12 level. The project was approved based on its potential benefit to society and the research community at large. The study was designed to minimize harm to the K-12 education community. We believe our contributions have sufficient novelty and utility to merit publication, and that the research poses minimal harm to those involved.


\section*{Open Science}
All artifacts necessary to replicate and evaluate this research are available in our \href{https://osf.io/hwkuc/overview?view_only=c226130f34fb4c28be314426390d6656}{supplementary materials}. These materials include the complete codebook with coding definitions and representative examples, the full database of analyzed and labeled standards, and the code used to execute the same database queries and analyses presented in the paper.

\bibliographystyle{plainurl}
\bibliography{teaching_standards}

\end{document}